\documentstyle[aps,prl,epsf]{revtex}
\begin{document}
\title{Normal modes and stability of phase separated trapped Bose-Einstein
condensates}
\date{\today }
\author{Anatoly A. Svidzinsky and Siu-Tat Chui}
\address{Bartol Research Institute, University of Delaware, Newark, DE 19716}
\maketitle

\begin{abstract}
We study the normal modes and the stability of two component condensates in
a phase separated regime. In such a regime the system can undergo a quantum
phase transition upon the change of interaction strength between bosons of
the same species or the variation of the trap frequencies. In this
transition, the distribution of the two components changes from a symmetric
to an asymmetric shape. We discuss the nature of the phase transition, the
role of the interface tension and the phase diagram. The symmetric to
asymmetric transition is the simplest quantum phase transition that one can
imagine. We found new branches of normal modes which are specific for
trapped multicomponent condensates and are analogous to the waves at the
interface between two layers of immiscible fluids under gravity. At the
point of the phase transition the frequencies of those modes go to zero and
become imaginary which causes an instability. The interface tension shifts
the normal mode frequencies and changes the stability region of the system.
\end{abstract}


\section{Introduction}

There is much recent interest in Bose-Einstein condensation (BEC) in trapped
gases as well as quantum phase transitions (QPT). Examples of the QPT
include the Wigner electron solid melting transition, the Mott-Hubbard
metal-insulator transition, and different magnetic transitions. In physical
phenomena involving BECs quantum mechanics play a crucial role. In this
paper we propose a new kind of quantum phase transition in phase-separated
mixtures of BECs. In this transition, the distribution of the two components
changes from a symmetric to an asymmetric shape. To explore the transition,
we first investigate the stability of the symmetric phase by studying its
normal modes. We find interface modes that become soft. These unstable modes
can be caused by changing the trapping potential and are analogous to
gravitational waves at the interface between two immiscible fluids. When the
lowest frequency becomes zero, the instability sets in which determines the
stability limit of the symmetric phase. We determine the actual phase
boundary by comparing the energy between the symmetric phase and the
asymmetric phase and find that the actual phase boundary and the instability
boundary is not the same. This suggests that the transition is first order.
The system may be a good laboratory to study issues of quantum metastability
and tunnelling. The symmetric to asymmetric transition is the simplest
quantum phase transition that one can imagine. Careful study of this problem
should provide us new insight into this burgeoning field of discovery.

Systems of multicomponent condensates were first realized by the JILA group
in a magnetic trap in $^{87}$Rb \cite{Myat97,Hall98} and subsequently in an
optical trap in $^{23}$Na (spinor condensate) \cite{Stam98,Sten98}.
Experimentally at low fields, the spin exchange process can occur in an
optically trapped condensate, leading to spin domains~\cite{Sten98} with
metastable behavior~\cite{Mies99,Stam99b}. Binary condensates in two
hyperfine levels of $^{87}$Rb have been studied~\cite{Myat97,Matt98}, most
notably realizing a system of interpenetrating Bose fluids~\cite{Hall98},
measurements of phase dispersion~\cite{Hall98b}, and a vortex state in a
dilute-gas BEC~\cite{Matt99a}. Theoretical treatment of such systems began
in the context of superfluid helium mixtures \cite{Khal57} and
spin-polarized hydrogen \cite{Sigg80}, and now has been extended to BEC in
trapped gases \cite
{Ho96,Law97,Esry97,Gold97,Ohbe98,Pu98,Ao98,Timm98,Chui99,Chui3}.

The equilibrium density distributions of segregated mixtures in the absence
of gravity have been studied numerically for different system parameters.
Two types of configurations have been discussed: a symmetric~\cite
{Pu98,Chui99,Esry97} configuration, for which one component is inside the
other one, and an asymmetric one in which the two components occupy the left
and the right hand side of a sphere~\cite{Chui99,Esry97}. Several
theoretical papers have described collective excitations of multicomponent
condensates. Busch {\it et al.} \cite{Busc97} studied collective excitations
in the limit of weak interactions based on a trial function approach.
Collective excitations of binary mixtures in the Thomas-Fermi (TF) limit
have been investigated by Graham and Walls \cite{Grah98}. Esry and Greene
\cite{Esry98} based on the Hartree-Fock and random-phase approximation have
numerically calculated the low-lying excitations of double BECs in a
time-averaged orbiting potential trap in which gravity separates the centers
of the two components. Gordon and Savage \cite{Gord98} have studied
excitation spectrum as a function of number of particles. Mazets \cite
{Maze02} has performed an analysis of wave dispersion on a boundary between
two weakly segregated untrapped BECs.

Under a change of the interaction strength the two component BECs can
undergo different types of macroscopic quantum phase transitions. For
example, an increase in interspecies interactions $a_{12}$ results in a
transition from a binary mixture to a phase separated state. Such phase
separated condensates are analogous to a system of immiscible fluids with
positive surface tension. Pu and Bigelow \cite{Pu98a} numerically studied
the frequency of collective excitations as a function of interspecies
interactions $a_{12}$ and found appearance of imaginary normal mode
frequencies at the phase separation transition point. Such imaginary modes
describe the nature of system's instability. The trapping potential is not
crucial for this type of phase transition and similar imaginary modes also
exist in the homogeneous (untrapped) BECs \cite{Gold97,Timm98,Grah98}.

However, in the phase separated regime the trapping potential changes the
system's symmetry and results in a new type of quantum phase transition.
Such transition occurs in trapped condensates upon the change of relative
intraspecies interaction strength $a_{11}/a_{22}$. When this ratio differs
significantly from one the less repulsive component is in the middle of the
trap and the more repulsive component comprises outer shell. One can make
the initial configuration unstable by manipulating the ratio $a_{11}/a_{22}$
close to one, e.g., by means of Feschbach resonances or by changing the
ratio of the trapping frequencies. In this paper we study the collective
excitations of the trapped phase separated condensates as a function of $%
a_{11}/a_{22}$ and show that normal modes with imaginary frequencies appear
when this ratio approaches one. Effect of the interface tension results in a
shift of the transition point of the symmetric-asymmetric transition in the
two component BEC. In general, the asymmetric phase possesses a lower
interface energy. On the other hand, since the degree of self-repulsion may
differ between the two species, the less self-repulsive component will
prefer to remain where the density is higher, while the other component
moves to the low density regions outside. This favors the symmetric phase.
Depending on the system parameters, one of these two energetic
considerations will win out. These system parameters can be adjusted by
changing the trapping frequencies, the relative particle numbers of the two
species, and the interaction between the particles with Feschbach
resonances. We first address the stability of the symmetric phase.

\section{Basic formalism}

Let us consider the two-component BEC in a spherically symmetric trap. The
dynamics of the system is described by time dependent Gross-Pitaevskii
equations

\begin{equation}
\label{a1}i\hbar \frac{\partial \Psi _1}{\partial t}=-\frac{\hbar ^2}{2m}%
\Delta \Psi _1+V_{{\rm tr}}\Psi _1+\frac{4\pi \hbar ^2}m(a_{11}|\Psi
_1|^2+a_{12}|\Psi _2|^2)\Psi _1,
\end{equation}
\begin{equation}
\label{a2}i\hbar \frac{\partial \Psi _2}{\partial t}=-\frac{\hbar ^2}{2m}%
\Delta \Psi _2+V_{{\rm tr}}\Psi _2+\frac{4\pi \hbar ^2}m(a_{22}|\Psi
_2|^2+a_{12}|\Psi _1|^2)\Psi _2,
\end{equation}
where $\Psi _{1,2}$ are the condensate wave functions, $V_{{\rm tr}}=m\omega
_0^2r^2/2$ is the trapping potential, $\omega _0$ is the trapping frequency,
$r$ is the radial spherical coordinate, $a_{ij}>0$ are $s-$wave scattering
lengths. Here we suppose that trapping frequencies and particle masses are
the same for both species. We discuss the general case in the last section.

We shall assume in this paper that the condition
\begin{equation}
\label{a21}a_{12}^2-a_{11}a_{22}>0
\end{equation}
is satisfied and, therefore, the condensates are phase-segregated. The first
demonstration of a condensate binary mixture by Myatt et. al. \cite{Myat97}
produced overlapping condensates of the $|F=1,m_f=-1>$ and $|F=2,m_f=2>$
spin states of $^{87}$Rb. These states, however, possess different magnetic
moments and, hence, the condensates experience different potentials in a
magnetic trap which results in unequal displacement from the trap center by
gravity. Later JILA\ experiments were performed on mixtures of $|1,-1>$ and $%
|2,1>$ states \cite{Hall98}. These two states have essentially identical
magnetic moments, and feel identical confining potentials. In $^{87}$Rb, the
scattering lengths for $|1,-1>$ and $|2,1>$ states are known to be in the
proportion $a_{11}:a_{12}:a_{22}::1.03:1:0.97$, with the average of the
three being $55(3)$\AA\ \cite{Matt98}. Hence, the condition (\ref{a21}) is
valid for the JILA experiments and the condensates are in a weakly
segregated phase \cite{Ao98}. One should mention that for earlier
experiments with the $|1,-1>$ and $|2,2>$ rubidium states the scattering
lengths are $a_{11}:a_{12}:a_{22}::1.007:1:1.01$ \cite{Burk97}. For such
states the condition (\ref{a21}) is not fulfilled and the mixture exhibits
behavior of a miscible system.

We suppose the condensates can be well described by the TF approximation. In
this regime, the phase-segregated condensates overlap over the length scale $%
\Lambda =\xi /\sqrt{a_{12}/\sqrt{a_{11}a_{22}}-1}$, where $\xi $ is the
healing length \cite{Ao98}. For the JILA experiments on phase-segregated
states $\Lambda \approx 47\xi $. If, however, the penetration depth $\Lambda
\ll R$, where $R $ is the size of the system, the condensates can be
approximately treated as nonoverlapping, which we assume to be the case. The
effect of the overlapping results in finite surface tension and can be
included via boundary conditions at the interface. However, in the TF limit
such effect is small, we shall discuss it in Sec. IV. If the condensates do
not overlap one can neglect the last terms in Eqs. (\ref{a1}), (\ref{a2}).
As a result, the dynamical equations for $\Psi _1$ and $\Psi _2$ decouple.
However, the two condensate components are coupled by boundary conditions at
the interface which require continuity of pressure and the normal velocity.

For definiteness we assume that stationary configuration is spherically
symmetric with the central core dominated by the first component and an
outer shell from the second species (see Fig. 1). The stationary density
distribution $n_i=|\Psi _i|^2$ of two components is given by
\begin{equation}
\label{d1}n_1=\frac{\mu _1}{G_{11}}\left( 1-\frac{r^2}{R_1^2}\right) ,\quad
0<r<R_{*},
\end{equation}
\begin{equation}
\label{d2}n_2=\frac{\mu _2}{G_{22}}\left( 1-\frac{r^2}{R_2^2}\right) ,\quad
R_{*}<r<R_2,
\end{equation}
where $G_{ii}=4\pi \hbar ^2a_{ii}/m$, $R_i=\sqrt{2\mu _i/m\omega _0^2}$. The
normalization condition $\int n_idV=N_i$, where $N_i$ are the numbers of
condensate particles, determines the chemical potentials $\mu _i$. The
position of the phase boundary $R_{*}$ is given by the condition that
pressures exerted by both condensates are equal \cite{Chui02}:
\begin{equation}
\label{a3}R_{*}=R_2\sqrt{\frac{1-\kappa \lambda }{(1-\kappa )\lambda }},
\end{equation}
where we have introduced dimensionless parameters
\begin{equation}
\label{a30}\kappa =\sqrt{\frac{a_{11}}{a_{22}}},\qquad \lambda =\frac{\mu _2%
}{\mu _1}.
\end{equation}

%%%%%%%%%%%%%%%%%%%% Fig 1 %%%%%%%%%%%%%%%%%%%%%%%%%%%%%
\begin{figure}
\bigskip
\centerline{\epsfxsize=0.53\textwidth\epsfysize=0.3\textwidth
\epsfbox{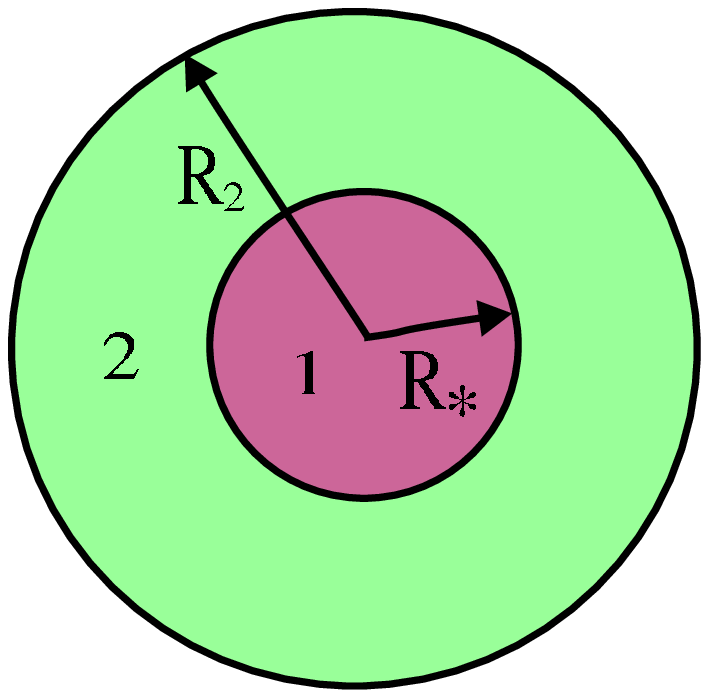}}

\label{fig1}
\end{figure}
%%%%%%%%%%%%%%%%%%%%%%%%%%%%%%%%%%%%%%%%%%%%%%%%%%%%%%%%%
\begin{center}
{\small Fig. 1. Symmetric configuration of BEC components.}
\end{center}

The symmetric configuration is favorable when $\kappa $ differs from unity,
with the less repulsive component being in the middle ($a_{11}<a_{22}$, that
is $\kappa <1$). At the interface between the two components the stationary
condensate densities are
\begin{equation}
\label{a31}n_1=\frac{\kappa (\mu _2-\mu _1)}{G_{11}(1-\kappa )},\quad n_2=%
\frac{(\mu _2-\mu _1)}{G_{22}(1-\kappa )}.
\end{equation}
$n_2/n_1=\kappa <1$.

One can rewrite the time dependent Gross-Pitaevskii equations in a
hydrodynamic form which shows an analogy between our problem and the motion
of two immiscible fluids. In the strong phase-segregated regime the dynamics
of each components is described by the following hydrodynamic equations

\begin{equation}
\label{s1}\frac{\partial n}{\partial t}+\nabla \cdot (n{\bf V})=0,
\end{equation}
\begin{equation}
\label{s2}\frac 12mV^2+V_{{\rm tr}}-\frac{\hbar ^2}{2m}\frac 1{\sqrt{n}}%
\Delta \sqrt{n}+Gn+m\frac{\partial \Phi }{\partial t}=\mu ,
\end{equation}
where ${\bf V}$ is the condensate velocity and $\Phi $ is the velocity
potential, ${\bf V=\nabla }\Phi $. Eq. (\ref{s1}) is continuity equation for
compressible flow, while Eq. (\ref{s2}) is the analog of the Bernoulli
equation with $(e+P)/n=-\frac{\hbar ^2}{2m}\frac 1{\sqrt{n}}\Delta \sqrt{n}%
+Gn$, where $e$ is the internal energy density and $P$ is the pressure. The
kinetic energy (quantum) pressure is omitted in the TF limit and, therefore,
$e=Gn^2/2$. As a result, from Eq. (\ref{s2}) we obtain for the pressure
\begin{equation}
\label{s3}2P=Gn^2=\mu n-nV_{{\rm tr}}-mn\frac{\partial \Phi }{\partial t}-%
\frac 12mnV^2.
\end{equation}
The trapping potential plays the role of gravitational potential in
hydrodynamics. The linearized hydrodynamic equations determine the
condensate normal modes. In terms of perturbation in the condensate density $%
n^{\prime }$ and the velocity potential $\Phi $ the normal-mode amplitudes
satisfy the coupled equations \cite{Fett96}
\begin{equation}
\label{a4}i\omega n^{\prime }=\nabla \cdot ({\bf V}n^{\prime })+\nabla \cdot
(n\nabla \Phi ),
\end{equation}
\begin{equation}
\label{a5}i\omega \Phi ={\bf V}\cdot \nabla \Phi +\frac Gmn^{\prime }-\frac{%
\hbar ^2}{4m^2n}\nabla \cdot \left[ n\nabla \left( \frac{n^{\prime }}n%
\right) \right] ,
\end{equation}
where $n$ is the static condensate density and ${\bf V}$ is the static
condensate velocity. In our problem ${\bf V=0}$, hence, $n^{\prime }\approx
i\omega m\Phi /G$ and the equation for the perturbation in velocity
potential is \cite{Stri96}
\begin{equation}
\label{e2}\frac{2\omega ^2}{\omega _0^2}\Phi -2r\frac{\partial \Phi }{%
\partial r}+(R^2-r^2)\Delta \Phi =0.
\end{equation}
One can seek solution of this equation in the form
$$
\Phi =\Phi (r)Y_{lm}(\theta ,\phi ),
$$
where $Y_{lm}(\theta ,\phi )$ is the spherical harmonics. Then, using $%
\Delta Y_{lm}(\theta ,\phi )=-l(l+1)/r^2$, we obtain the following equation
for the radial function $\Phi (r)$
\begin{equation}
\label{e3}r^2(r^2-R^2)\frac{d^2\Phi }{dr^2}+(4r^2-2R^2)r\frac{d\Phi }{dr}%
+\left[ l(l+1)R^2-r^2\left( \frac{2\omega ^2}{\omega _0^2}+l(l+1)\right)
\right] \Phi =0.
\end{equation}
Let us derive boundary conditions for $\Phi $ at the interface. The change
in density at the interface is the sum of the term from the movement of the
interface and that from the perturbation of the density. Let $\varsigma
=\varsigma (\theta ,\phi ,t)$ be a small deviation of the $r-$coordinate of
the interphase from its stationary value $R_{*}$. Then the condensate
density at the interface is $n(R_{*}+\varsigma ,t)\approx $ $n(R_{*},t)+$ $%
\varsigma \partial n(R_{*},t)/\partial r$, and, hence, the pressure is given
by
\begin{equation}
\label{s7}P=\frac{Gn^2}2\approx \frac{Gn^2(R_{*},t)}2+Gn(R_{*},t)\frac{%
\partial n(R_{*},t)}{\partial r}\varsigma .
\end{equation}
In terms of the density perturbation $n^{\prime }$, $n(R_{*},t)=n(R_{*})+n^{%
\prime }(R_{*},t)$, we obtain
\begin{equation}
\label{s8t}P\approx \frac{Gn^2(R_{*})}2+Gn(R_{*})n^{\prime
}(R_{*},t)-n(R_{*})m\omega _0^2R_{*}\varsigma .
\end{equation}
Continuity of the pressure at the interface results in the following
boundary condition%
$$
G_{11}n_1(R_{*})n_1^{\prime }(R_{*},t)-n_1(R_{*})m\omega _0^2R_{*}\varsigma
_1=G_{22}n_2(R_{*})n_2^{\prime }(R_{*},t)-n_2(R_{*})m\omega
_0^2R_{*}\varsigma _2.
$$
Taking into account $Gn^{\prime }=im\omega \Phi $ and $\partial \varsigma
/\partial t=V_r=\partial \Phi /\partial r$ we finally obtain the following
boundary condition for the velocity potential at the interface
\begin{equation}
\label{s10}\omega _0^2R_{*}\frac{\partial \Phi _1}{\partial r}-\omega ^2\Phi
_1=\kappa \omega _0^2R_{*}\frac{\partial \Phi _2}{\partial r}-\kappa \omega
^2\Phi _2.
\end{equation}
Continuity of $V_r$ gives another boundary condition
\begin{equation}
\label{s11}\frac{\partial \Phi _1}{\partial r}=\frac{\partial \Phi _2}{%
\partial r}.
\end{equation}
Eq. (\ref{e3}) and the boundary conditions (\ref{s10}), (\ref{s11}) compose
a complete set of equations necessary to determine normal modes of the
system.

\section{Normal modes}

For one component condensates the solutions of Eq. (\ref{e3}) which are
finite at $r=0$ and $r=R$ are
\begin{equation}
\label{e31}\Phi (r)\propto r^lP_n^{(l+\frac 12,0)}\left( 1-\frac{2r^2}{R^2}%
\right) ,
\end{equation}
where $P_n^{(l+\frac 12,0)}(x)$ are Jacobi polynomials, $n$ is the radial
quantum number. The corresponding eigenfrequencies are given by \cite{Stri96}
\begin{equation}
\label{e32}\omega ^2=\omega _0^2(l+n(2n+2l+3)).
\end{equation}
For the two component condensate the normal mode frequencies are different
from the general case and new modes appear. However, we shall show that
frequencies corresponding to $n=0$ remain the same as for the one component
system. Rewriting the function $\Phi =r^l\eta (r^2/R^2)$ in Eq. (\ref{e3})
results in a hypergeometric equation for $\eta (\xi )$%
\begin{equation}
\label{e4}\xi (\xi -1)\eta _{\xi \xi }^{\prime \prime }+\left[ \left( l+%
\frac 52\right) \xi -l-\frac 32\right] \eta _\xi ^{\prime }+\frac 12\left( l-%
\frac{\omega ^2}{\omega _0^2}\right) \eta =0.
\end{equation}
Hence, general solutions for the inner and outer condensates which are
regular at $r=0$ and $r=R_2$ respectively have the form
\begin{equation}
\label{e6}\Phi _1=C_1r^lF\left( \alpha ,\beta ,l+3/2,\frac{r^2}{R_1^2}%
\right) ,\quad \Phi _2=C_2r^lF\left( \alpha ,\beta ,1,1-\frac{r^2}{R_2^2}%
\right) ,
\end{equation}
where $F$ is hypergeometric function, $C_1$, $C_2$ are constants and
$$
\alpha =\frac 12\left[ l+\frac 32-\sqrt{l^2+l+\frac 94+\frac{2\omega ^2}{%
\omega _0^2}}\right] ,\quad \beta =\frac 12\left[ l+\frac 32+\sqrt{l^2+l+%
\frac 94+\frac{2\omega ^2}{\omega _0^2}}\right] .
$$
Using Eq. (\ref{e6}), the boundary conditions (\ref{s10}), (\ref{s11}) and
the mathematical identity
$$
\frac d{dz}F(\alpha ,\beta ,\gamma ,z)=\frac{\alpha \beta }\gamma F(\alpha
+1,\beta +1,\gamma +1,z),
$$
we obtain the equation for the normal mode frequencies
\begin{equation}
\label{s16}\frac{\omega ^2}{\omega _0^2}=(1-\kappa )\frac{\left[
l(l+3/2)+(l-\omega ^2/\omega _0^2)\lambda xs_1(\omega ,x)\right] \left[
(l-\omega ^2/\omega _0^2)xs_2(\omega ,x)-l\right] }{\left[ l(l+3/2)(\kappa
-1)+x(l-\omega ^2/\omega _0^2)\left( \kappa \lambda s_1(\omega
,x)+(l+3/2)s_2(\omega ,x)\right) \right] },
\end{equation}
where%
$$
s_1(\omega ,x)=\frac{F\left( \alpha +1,\beta +1,l+5/2,\lambda x\right) }{%
F\left( \alpha ,\beta ,l+3/2,\lambda x\right) },\quad s_2(\omega ,x)=\frac{%
F\left( \alpha +1,\beta +1,2,1-x\right) }{F\left( \alpha ,\beta
,1,1-x\right) },\quad x=\frac{R_{*}^2}{R_2^2}.
$$
One of the solutions of Eq. (\ref{s16}) is $\omega ^2=l\omega _0^2$, which
coincide with those for one component condensate. For this solution the
components oscillate in-phase and $\Phi _1=\kappa \Phi _2\propto
r^lY_{lm}(\theta ,\phi )$. Another exact solution is $\omega ^2=5\omega _0^2$
with $\Phi _1\propto 1-5r^2/3R_1^2$, $\Phi _2\propto \lambda (1-5r^2/3R_2^2)$
which corresponds to $l=0$, $n=1$. For this solution the components
oscillate out-of-phase if $\sqrt{3/5}R_1<R_{*}<\sqrt{3/5}R_2$ and in-phase
otherwise.

\subsection{Numerical results}

We solve Eq. (\ref{s16}) numerically and find normal mode frequencies $%
\omega $ of the two component condensate as a function of the parameter $%
\kappa =\sqrt{a_{11}/a_{22}}$ for different fixed ratios $R_{*}/R_2$. The
ratio $R_{*}/R_2$ can be directly measured experimentally. Fig. 2 shows the
normal mode frequencies for $l=0$. At fixed $l$ there is an infinite number
of branches which correspond to different radial quantum numbers $n$. We
plot only the lowest modes with $n=0,1,2$. In Fig. 3 the normal modes are
estimated for $l=1$. In the limiting case $a_{11}=a_{22}$ ($\kappa =1$) the
condensates behave as an one component system and the condensate modes
coincide with Stringari's result (\ref{e32}). Fig. 4 shows the low frequency
modes that become imaginary at $\kappa >1$. These modes are peculiar for two
component systems and are analogous to the waves at the interface between
two layers of immiscible fluids under gravity \cite{Land88a}. In trapped
condensates the gradient of trap potential plays the role of gravitational
field. As soon as $\kappa $ becomes greater than $1$, which means $n_2>n_1$
at the interface, the system becomes unstable the same way as two immiscible
fluids in gravitational field when the more dense layer is on the top. One
should mention that normal modes with imaginary frequency also appear at the
transition point from a binary mixture into a phase separated state which
occurs at $a_{12}=\sqrt{a_{11}a_{22}}$ \cite{Timm98,Grah98,Pu98a}. In that
case, the fastest decaying mode occurs at a finite wavevector, resulting in
a quantum spinodal transition \cite{qsd}.

%%%%%%%%%%%%%%%%%%%% Fig 2 %%%%%%%%%%%%%%%%%%%%%%%%%%%%%
\begin{figure}
\bigskip
\centerline{\epsfxsize=0.48\textwidth\epsfysize=0.55\textwidth
\epsfbox{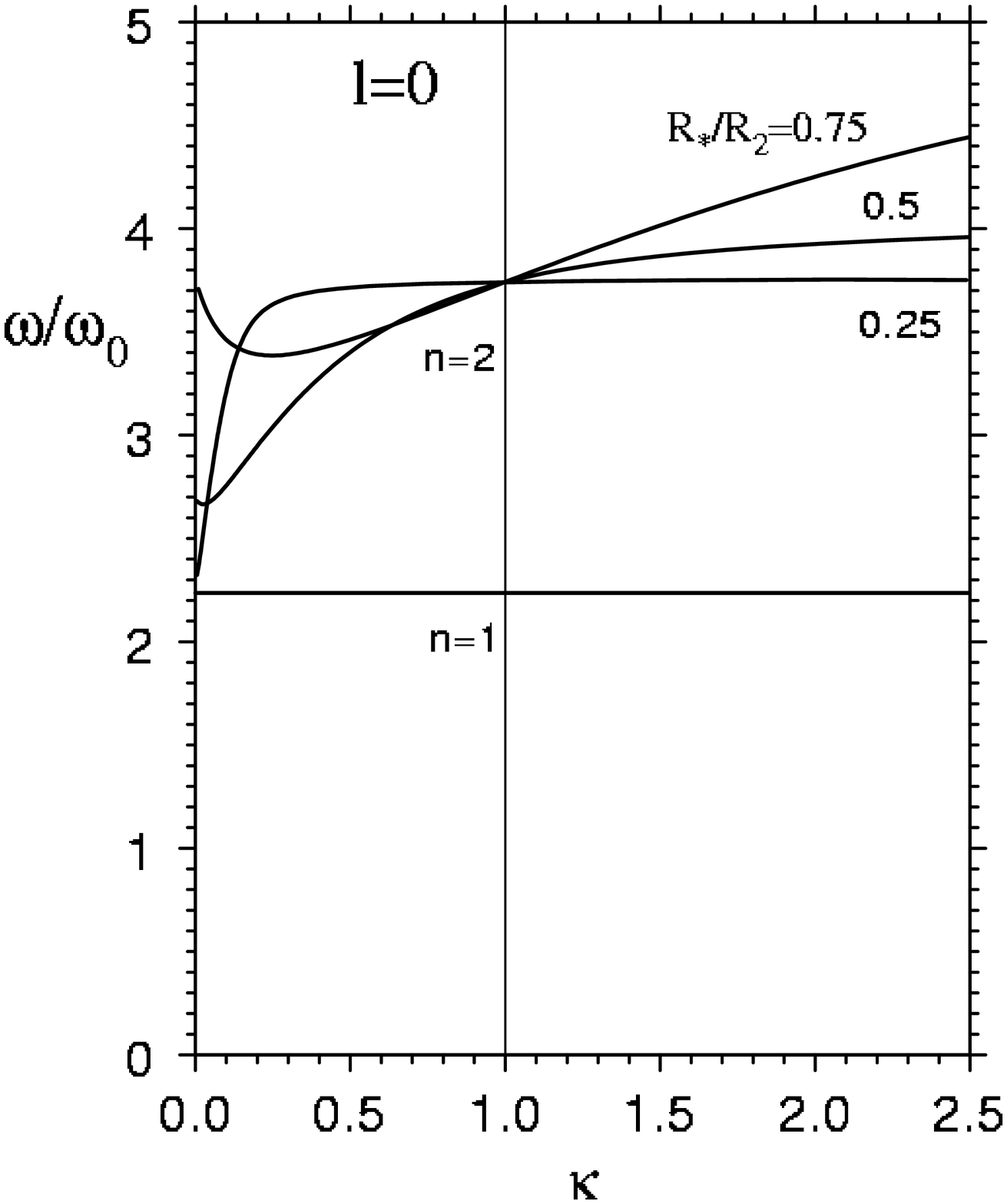}}

\label{fig2}
\end{figure}
%%%%%%%%%%%%%%%%%%%%%%%%%%%%%%%%%%%%%%%%%%%%%%%%%%%%%%%%%

{\small Fig. 2. Normal mode frequencies for $l=0$ and $n=1,2$ as a function
of $\kappa =\sqrt{a_{11}/a_{22}}$. The mode with $n=2$ depends on position of
the interface, the mode frequency is estimated for $R_{*}/R_2=0.25,0.5,0.75$.
For $\kappa =1$ the mode frequencies coincide with those for the one
component condensate.}

%%%%%%%%%%%%%%%%%%%% Fig 3 %%%%%%%%%%%%%%%%%%%%%%%%%%%%%
\begin{figure}
\bigskip
\centerline{\epsfxsize=0.48\textwidth\epsfysize=0.48\textwidth
\epsfbox{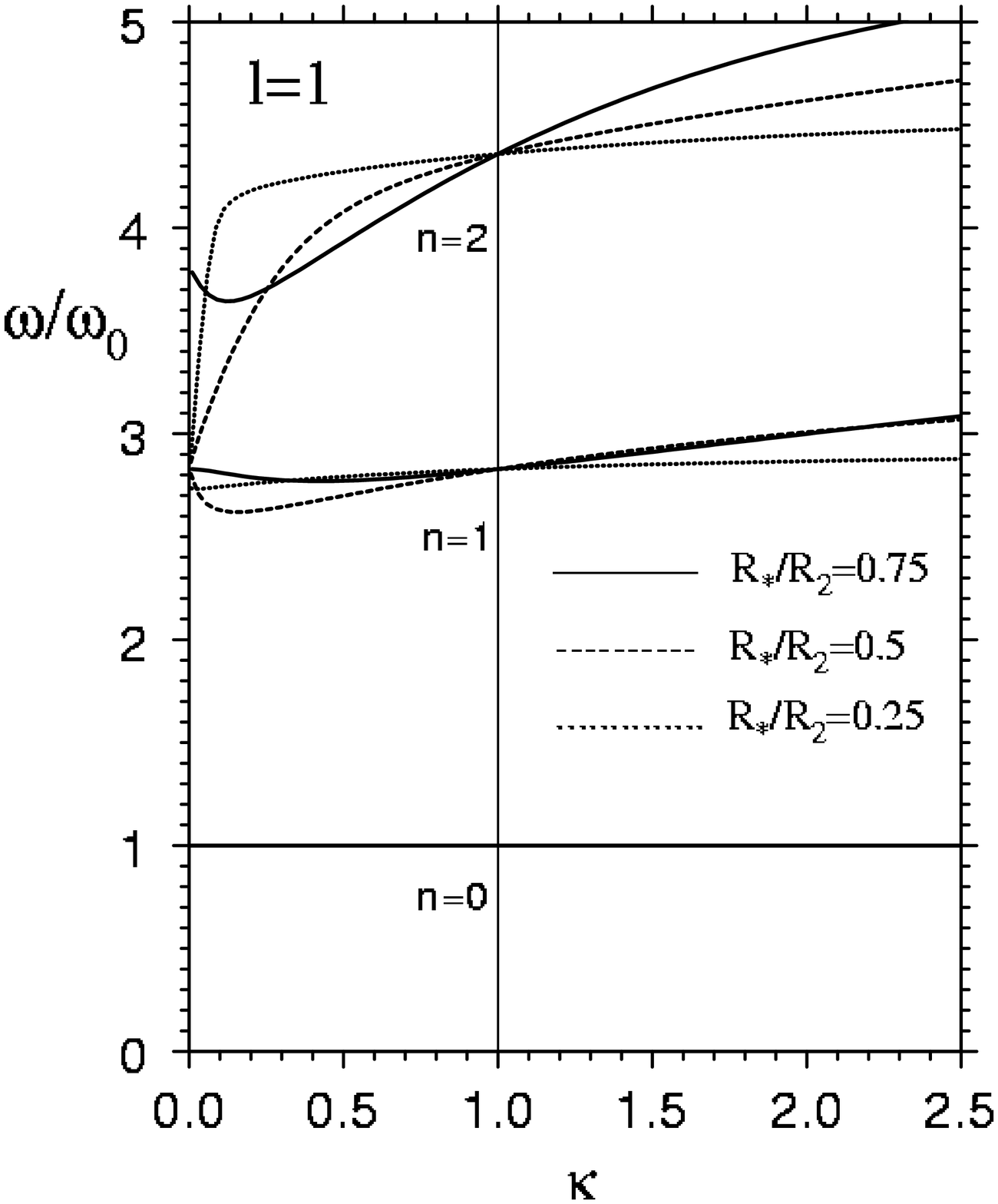}}

\label{fig3}
\end{figure}
%%%%%%%%%%%%%%%%%%%%%%%%%%%%%%%%%%%%%%%%%%%%%%%%%%%%%%%%%

\begin{center}
{\small Fig. 3. Normal mode frequencies for $l=1$ and $n=0,1,2$.}
\end{center}

%%%%%%%%%%%%%%%%%%%% Fig 4 %%%%%%%%%%%%%%%%%%%%%%%%%%%%%
\begin{figure}
\bigskip
\centerline{\epsfxsize=0.48\textwidth\epsfysize=0.48\textwidth
\epsfbox{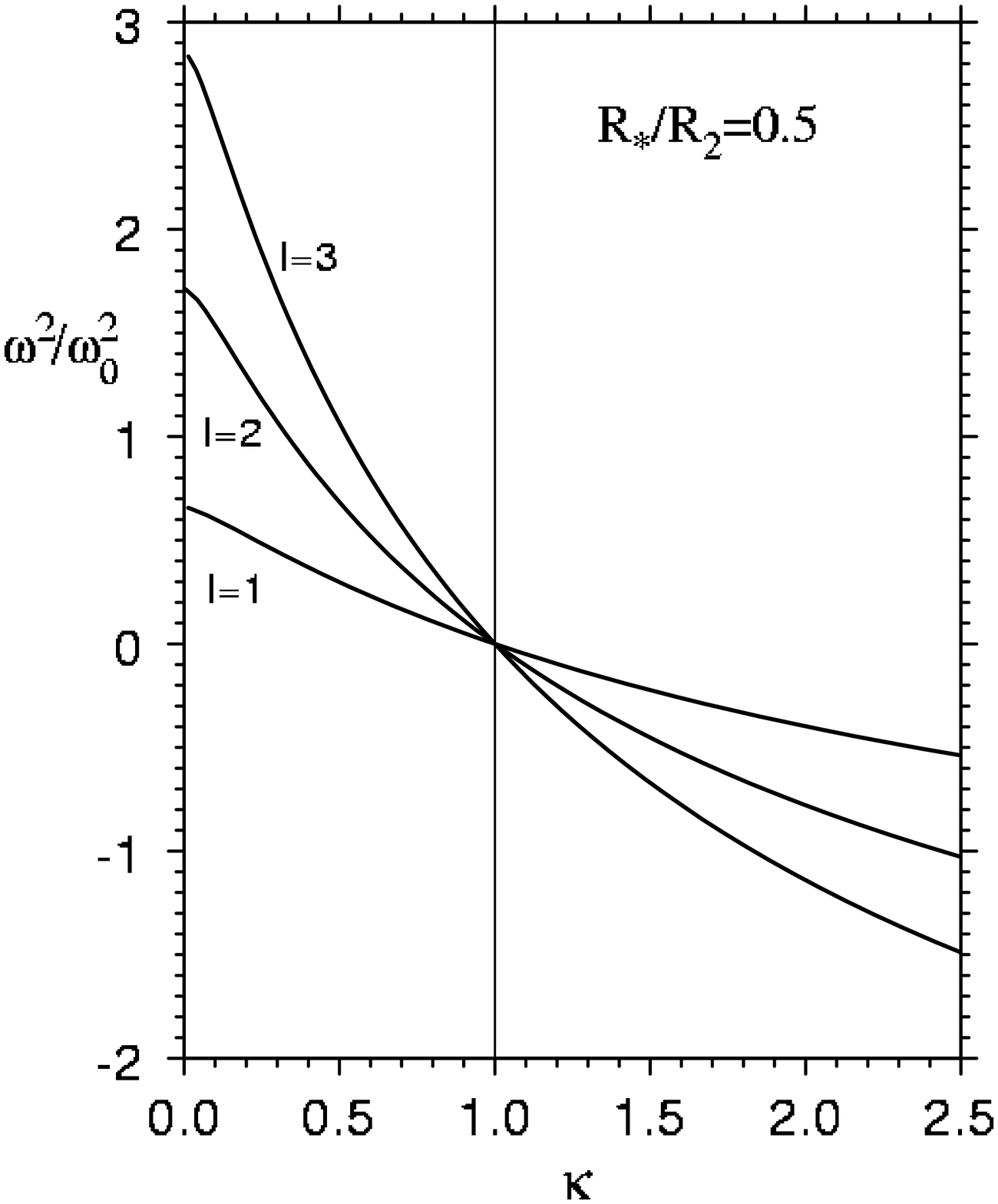}}

\label{fig4}
\end{figure}
%%%%%%%%%%%%%%%%%%%%%%%%%%%%%%%%%%%%%%%%%%%%%%%%%%%%%%%%%

{\small Fig. 4. Low frequency modes as a function of $\kappa
=\sqrt{a_{11}/a_{22}}$ for different $l=1,2,3$. The position of the interface
is $R_{*}=R_2/2$. Frequencies become imaginary at $\kappa >1$.}

\subsection{Low frequency modes}

Here we study the low frequency modes in detail. In the region $|1-\kappa
|\ll 1\,$the mode frequencies are small: $|\omega |\ll \omega _0$. In this
limit one can take $\omega =0$ in the right side of Eq. (\ref{s16}) and put $%
\kappa ,\lambda \approx 1$ in the multiple. As a result, we obtain
\begin{equation}
\label{s19}\omega ^2\approx \omega _0^2(1-\kappa )f(l,x),
\end{equation}
where%
$$
f(l,x)=\frac{l\left( l+3/2+xs_1(0,x)\right) \left( xs_2(0,x)-1\right) }{%
x\left[ s_1(0,x)+(l+3/2)s_2(0,x)\right] }.
$$
Eq. (\ref{s19}) describes behavior of the low frequency modes in the region
close to the point of instability $\kappa =1$. In this region $\omega
\propto \sqrt{1-\kappa }$ and becomes imaginary when $\kappa >1$. In Fig. 5
we plot $f$ as a function of $x=R_{*}/R_2$ for different $l$. Imaginary part
of frequencies is greater for larger $l$ and decreases with increasing $%
R_{*}/R_2$, that is bigger inner droplets are more stable. Figs. 6, 7 show
the radial distribution of the density perturbation for the low frequency
modes with different $l$. In our estimates we choose the radius of the inner
droplet $R_{*}=R_2/2$. The density perturbations are normalized so that $%
n_2^{\prime }(R_2)=1$. For the low frequency modes the two condensates
oscillate out-of-phase. Fig. 6 corresponds to $\kappa =0.9$ (stable region),
while for Fig. 7 $\kappa =1.1$ (unstable regime). One can see that the
normal mode profile undergoes no changes at the point of phase transition,
although the mode frequencies become imaginary at this point. However, due
to the relation $n^{\prime }\approx i\omega m\Phi /G$, the phase shift
between oscillations of $n^{\prime }$ and $\Phi $ changes at the instability
point from $\pi /2$ to $0$. Modes with small $l$ are delocalized and the
whole condensate is involved in oscillations. With increasing $l$, however,
the modes become localized near the interface and in the limit $l\rightarrow
\infty $ they are similar to gravitational waves at the surface of deep
water with the dispersion relation $\omega \propto \sqrt{k}$.

%%%%%%%%%%%%%%%%%%%% Fig 5 %%%%%%%%%%%%%%%%%%%%%%%%%%%%%
\begin{figure}
\bigskip
\centerline{\epsfxsize=0.48\textwidth\epsfysize=0.48\textwidth
\epsfbox{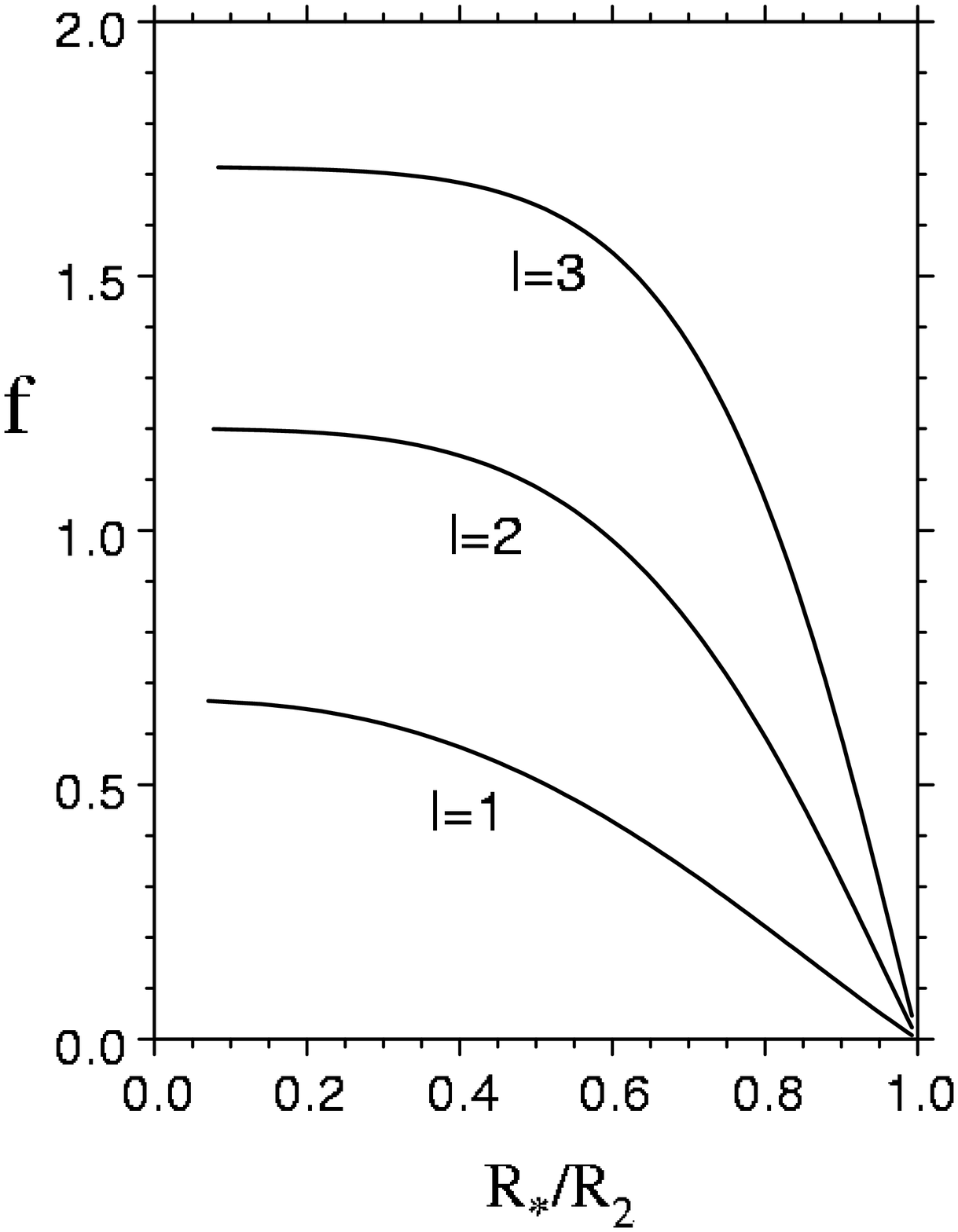}}

\label{fig5}
\end{figure}
%%%%%%%%%%%%%%%%%%%%%%%%%%%%%%%%%%%%%%%%%%%%%%%%%%%%%%%%%

{\small Fig. 5. Dependence of the factor $f$ in Eq. (\ref{s19}) on the position
of the interface $R_{*}/R_2$ for modes with different $l=1,2,3$.}

%%%%%%%%%%%%%%%%%%%% Fig 6 %%%%%%%%%%%%%%%%%%%%%%%%%%%%%
\begin{figure}
\bigskip
\centerline{\epsfxsize=0.48\textwidth\epsfysize=0.45\textwidth
\epsfbox{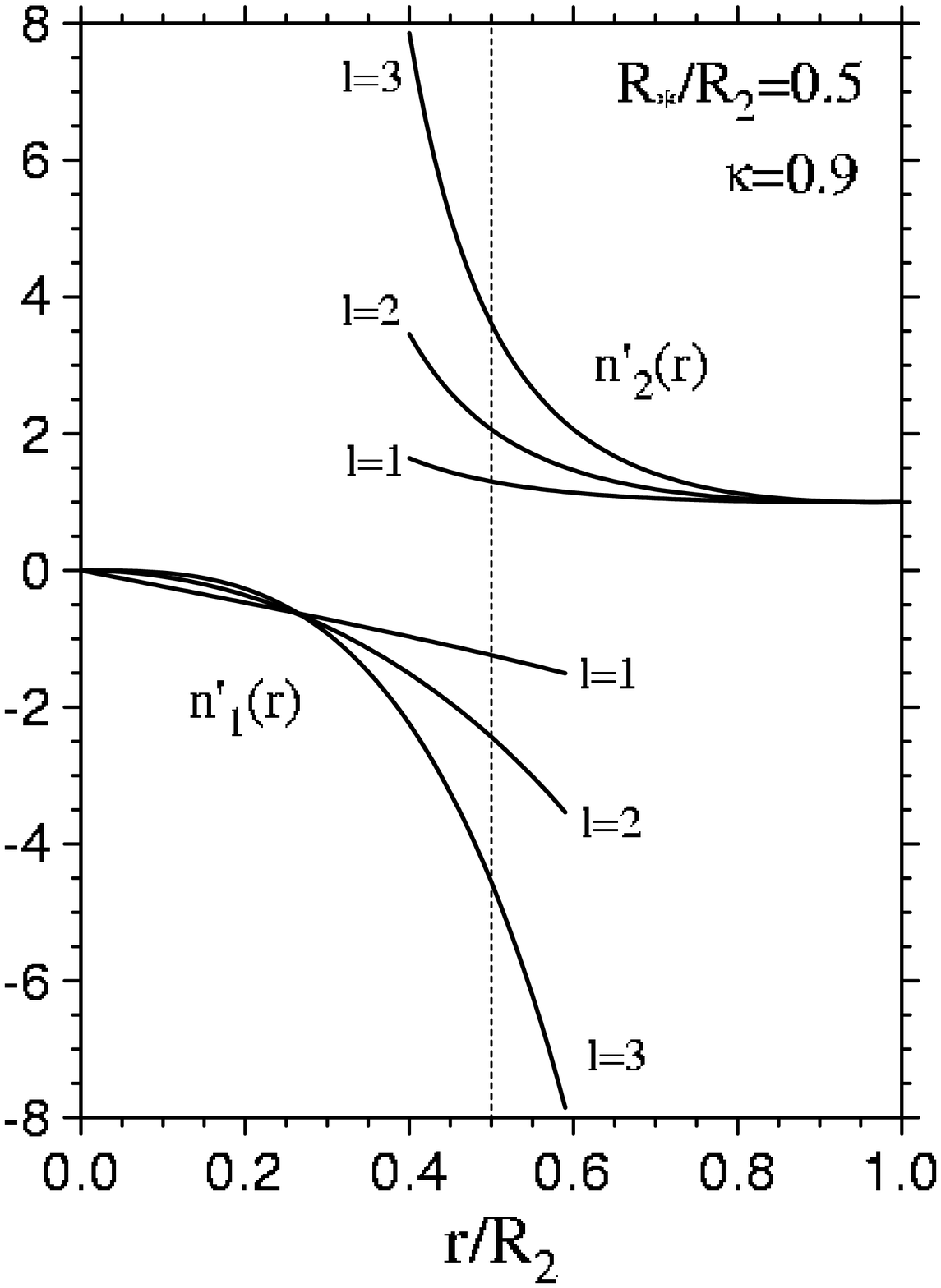}}

\label{fig6}
\end{figure}
%%%%%%%%%%%%%%%%%%%%%%%%%%%%%%%%%%%%%%%%%%%%%%%%%%%%%%%%%

{\small Fig. 6. Perturbation in the condensate density of the two
components $n_1^{\prime }(r)$ and $n_2^{\prime }(r)$ for the low frequency
modes with different $l=1,2,3$. The position of the interface is
$R_{*}=R_2/2$ (dash line) and $\kappa =0.9$ (stable region).}

%%%%%%%%%%%%%%%%%%%% Fig 7 %%%%%%%%%%%%%%%%%%%%%%%%%%%%%
\begin{figure}
\bigskip
\centerline{\epsfxsize=0.48\textwidth\epsfysize=0.45\textwidth
\epsfbox{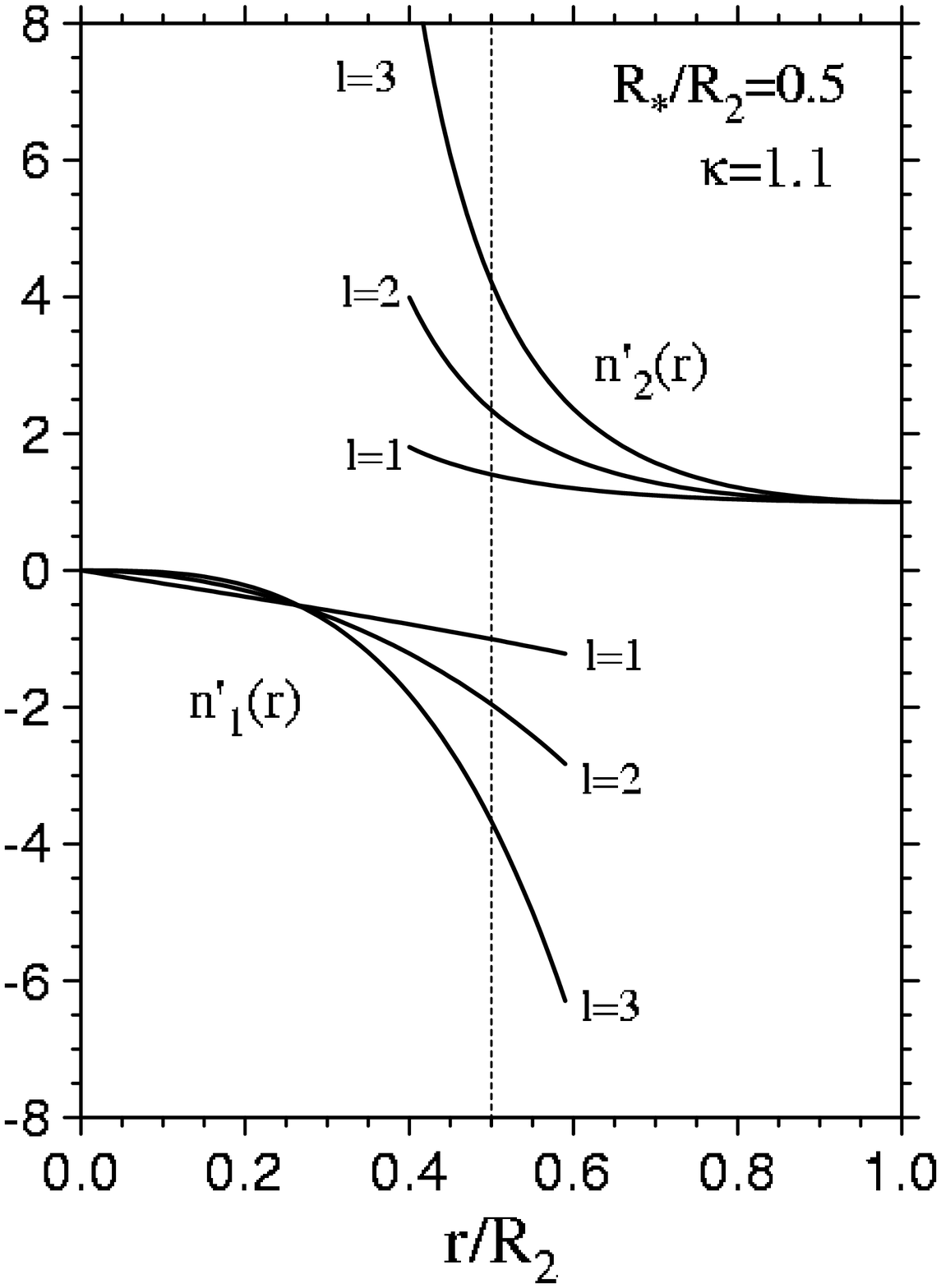}}

\label{fig7}
\end{figure}
%%%%%%%%%%%%%%%%%%%%%%%%%%%%%%%%%%%%%%%%%%%%%%%%%%%%%%%%%

\begin{center}
{\small Fig. 7. The same as in Fig. 6, but for $\kappa =1.1$ (unstable region).}
\end{center}

\section{Effect of interface tension}

Interface tension is small in the TF limit and results in corrections of the
order of $\xi /R_2$ to the normal mode frequencies. However, for the low
frequency modes in the vicinity $\kappa \approx 1$ the effect of the
interface tension is substantial because the mode frequencies themselves are
close to zero. The surface tension $\sigma $ modifies the boundary condition
for the pressure so that the pressure difference at the interface is equal
to the surface tension pressure $P_1-P_2=\sigma (1/r_1+1/r_2)$, where $r_1$,
$r_2$ are the principle radii of curvature. The stationary position of the
interface is determined by the equation%
$$
\frac{G_{11}n_1^2}2=\frac{G_{22}n_2^2}2+\frac{2\sigma }{R_{*}},
$$
which results in the following ratio of densities at the interface%
$$
\frac{n_2}{n_1}=\sqrt{\kappa ^2-\frac{4\sigma }{G_{22}n_1^2R_{*}}}=\kappa _{%
{\rm eff}}.
$$
The interface shape oscillates during the condensate motion which produces
oscillations of the surface pressure. For interface shape close to spherical
\cite{Land88}
\begin{equation}
\label{s20}\frac 1{r_1}+\frac 1{r_2}=\frac 2{R_{*}}-\frac{2\varsigma }{%
R_{*}^2}-\frac{\Delta _{\theta ,\phi }\varsigma }{R_{*}^2},
\end{equation}
where $\Delta _{\theta ,\phi }$ is the angular part of Laplace's operator.
Also one should take into account that $\sigma $ itself depends on local
density and varies during the interface motion. We assume $\sigma \propto
\sqrt{\frac 1{G_{11}n_1}+\frac 1{G_{22}n_2}}n_1n_2$ \cite{Timm98}. Then,
taking into account $n(R_{*}+\varsigma ,t)\approx $ $n(R_{*})+n^{\prime
}(R_{*},t)+$ $\varsigma \partial n(R_{*})/\partial r$, we obtain%
$$
\sigma =\sigma _0\left( 1+\frac m{G_{11}n_1}\left( i\omega \Phi _1-\omega
_0^2R_{*}\varsigma \right) \left[ \frac{1+2\kappa _{{\rm eff}}}{2(1+\kappa _{%
{\rm eff}})}\right] +\frac m{G_{22}n_2}\left( i\omega \Phi _2-\omega
_0^2R_{*}\varsigma \right) \left[ \frac{2+\kappa _{{\rm eff}}}{2(1+\kappa _{%
{\rm eff}})}\right] \right) ,
$$
where $\sigma _0$ is the tension for the stationary configuration. As a
result, the boundary condition (\ref{s10}) contains extra terms and becomes%
$$
\left( mn_1-\frac{2m\sigma _0}{R_{*}G_{11}n_1}\left[ \frac{1+2\kappa _{{\rm %
eff}}}{2(1+\kappa _{{\rm eff}})}\right] \right) \left[ \omega _0^2R_{*}\frac{%
\partial \Phi _1}{\partial r}-\omega ^2\Phi _1\right] -\frac{\sigma _0}{%
R_{*}^2}\left( 2\frac{\partial \Phi _1}{\partial r}+\Delta _{\theta ,\phi }%
\frac{\partial \Phi _1}{\partial r}\right) =
$$
\begin{equation}
\label{s21}=\left( mn_2+\frac{2m\sigma _0}{R_{*}G_{22}n_2}\left[ \frac{%
2+\kappa _{{\rm eff}}}{2(1+\kappa _{{\rm eff}})}\right] \right) \left[
\omega _0^2R_{*}\frac{\partial \Phi _2}{\partial r}-\omega ^2\Phi _2\right]
,
\end{equation}
while Eq. (\ref{s11}) remains the same. For the radial part of the velocity
potential we obtain%
$$
\left( 1-\frac{\sigma _0}{R_{*}P}\left[ \frac{1+2\kappa _{{\rm eff}}}{%
2(1+\kappa _{{\rm eff}})}\right] \right) \left[ \omega _0^2R_{*}\frac{%
\partial \Phi _1}{\partial r}-\omega ^2\Phi _1\right] +\frac{\sigma
_0(l-1)(l+2)}{mn_1R_{*}^2}\frac{\partial \Phi _1}{\partial r}=
$$
\begin{equation}
\label{s22}=\kappa _{{\rm eff}}\left( 1+\frac{\sigma _0}{R_{*}P}\left[ \frac{%
2+\kappa _{{\rm eff}}}{2(1+\kappa _{{\rm eff}})}\right] \right) \left[
\omega _0^2R_{*}\frac{\partial \Phi _2}{\partial r}-\omega ^2\Phi _2\right]
,
\end{equation}
where $P$ is the pressure at the interface. The modified boundary condition
results in the following equation for the low frequency modes
\begin{equation}
\label{s23}\omega ^2\approx \omega _0^2\left( 1-\frac{3\sigma _0}{2R_{*}P}+%
\frac{\sigma _0(l-1)(l+2)}{mn_1\omega _0^2R_{*}^3}-\kappa _{{\rm eff}%
}\right) f.
\end{equation}
where $n_1$ is the density at the interface. The interface tension shifts
the frequencies of the lowest modes and narrows the stability region. For $%
l=1$ the inner droplet moves as a whole without changing its shape. However,
the displacement of the droplet into the less dense region decreases the
interface energy which is proportional to $n^{3/2}$. This is the origin for
the contribution $-3\sigma _0/R_{*}P$ to the mode frequency. The mode
frequencies now become imaginary at
\begin{equation}
\label{s25}\kappa ^2>1-\frac{\sigma _0}{R_{*}P}+\frac{2\sigma _0(l-1)(l+2)}{%
mn_1\omega _0^2R_{*}^3}.
\end{equation}
As $\kappa $ increases the mode with $l=1$ becomes imaginary first which
determines the system's stability limit. When the surface tension is absent,
the higher $l$ is the more unstable the mode is. In the present case, the
two component condensate is locally unstable when
\begin{equation}
\label{s26}\frac{a_{11}}{a_{22}}>1-\frac{\sigma _0}{R_{*}P}.
\end{equation}

\section{Global stability: Energy of asymmetric phase}

So far we have discussed the {\bf local} stability of the symmetric phase.
We next turn our attention to its {\bf global} stability by comparing the
energies of the symmetric phase and the asymmetric phase. Let us consider an
asymmetric state in which component 1 sits on the top of the component 2.
The transition occurs when $\kappa $ is close to $1$ which we assume. Hence,
the interface is approximately flat (see Fig. 8).

%%%%%%%%%%%%%%%%%%%% Fig 8 %%%%%%%%%%%%%%%%%%%%%%%%%%%%%
\begin{figure}
\bigskip
\centerline{\epsfxsize=0.71\textwidth\epsfysize=0.4\textwidth
\epsfbox{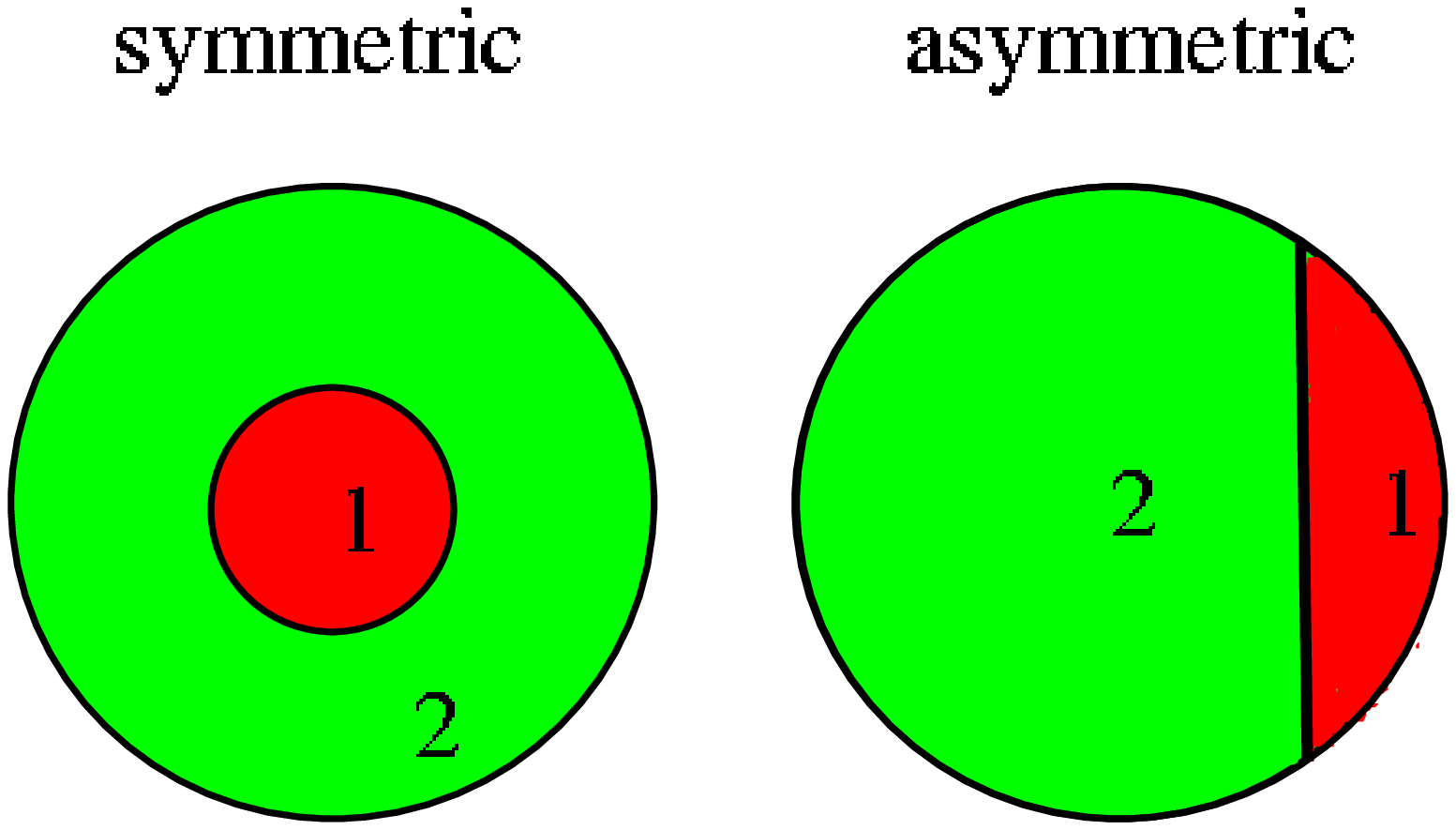}}

\label{fig8}
\end{figure}
%%%%%%%%%%%%%%%%%%%%%%%%%%%%%%%%%%%%%%%%%%%%%%%%%%%%%%%%%
\begin{center}
{\small Fig. 8. Symmetric and Asymmetric phases of two component BECs.}
\end{center}

In the Thomas-Fermi approximation, the density distribution is still given
by Eqs. (\ref{d1}) and (\ref{d2}), but the interface position is now
different. The TF density distribution suggests that the outer edge of the
condensates has a spherical shape and, therefore, in the asymmetric
configuration the condensates have equal chemical potentials $\mu _{1a}=\mu
_{2a}=\mu $. Carrying out the integration, we find that the energy of the
asymmetric state is given by
\begin{equation}
\label{b42}E_a=\frac \mu 2\left( N_1+N_2\right) +\frac{\pi m^2\omega _0^4R^7}%
8\left( \frac 1{G_{11}}-\frac 1{G_{22}}\right) \left[ \frac 4{35}-\frac d{6R}%
+\frac{d^5}{10R^5}-\frac{d^7}{21R^7}\right] +\frac{\pi m^2\omega _0^4R^7}{%
35G_{22}}+E_{sa},
\end{equation}
where $z=d$ is the position of the interface, $E_{sa}$ is the interface
energy and $R=\sqrt{2\mu /m\omega _0^2}$ is the radius of the sphere. The
radius $R$ (and hence $\mu $) and the position $d$ are determined by the
number of particles
\begin{equation}
\label{b43}N_1=\frac{\pi \mu R^3}{15G_{11}}\left[ 4-\frac{15d}{2R}+\frac{5d^3%
}{R^3}-\frac{3d^5}{2R^5}\right] ,
\end{equation}
\begin{equation}
\label{b44}N_2=\frac{\pi \mu R^3}{15G_{22}}\left[ 4+\frac{15d}{2R}-\frac{5d^3%
}{R^3}+\frac{3d^5}{2R^5}\right] .
\end{equation}
From Eqs. (\ref{b43}), (\ref{b44}), we obtain
\begin{equation}
\label{b45}G_{11}N_1+G_{22}N_2=\frac{4\pi m\omega _0^2R^5}{15},
\end{equation}
which determines $R$. The position of the interface can be found from
\begin{equation}
\label{b46}4-\frac{15d}{2R}+\frac{5d^3}{R^3}-\frac{3d^5}{2R^5}=\frac{%
8G_{11}N_1}{G_{22}N_2+G_{11}N_1}.
\end{equation}
The energy of a spherically symmetric configuration when the first droplet
is located inside and the second outside is given by
$$
E_0=\frac{\mu _1N_1}2+\frac{\mu _2N_2}2+\pi m\omega _0^2\left[ \left( \frac{%
\mu _1}{G_{11}}-\frac{\mu _2}{G_{22}}\right) \frac{R_{*}^5}5+\frac{m\omega
_0^2}2\left( \frac 1{G_{22}}-\frac 1{G_{11}}\right) \frac{R_{*}^7}7+\right.
$$
\begin{equation}
\label{b500}\left. +\frac{2^{7/2}\mu _2^{7/2}}{35G_{22}m^{5/2}\omega _0^5}%
\right] +E_{s0},
\end{equation}
where $E_{s0}$ is the interface energy in the symmetric state.

When the system is close to the point of phase transition the interface
energies of two configurations are given by (for details see Sec. VI)
\begin{equation}
\label{b491}E_{s0}=\frac{8\pi \hbar R_{*}^2\mu ^{3/2}}{\sqrt{3m}G}\sqrt{%
a_{12}/\sqrt{a_{11}a_{22}}-1}\left( 1-\frac{R_{*}^2}{R^2}\right) ^{3/2},
\end{equation}
\begin{equation}
\label{b492}E_{sa}=\frac{4\pi \hbar R^2\mu ^{3/2}}{5\sqrt{3m}G}\sqrt{a_{12}/%
\sqrt{a_{11}a_{22}}-1}\left( 1-\frac{d^2}{R^2}\right) ^{5/2}.
\end{equation}
In general, these expressions need to be numerically evaluated. They are
analytically tractable in the limit when one of the components is
particularly small, to which we turn our attention next.

\subsection{Energy for $N_1\ll N_2$}

If the number of atoms of the first species is much less than $N_2$ then $%
R_{*}\ll R$, $d/R\approx 1-(4N_1/5N_2)^{1/3}$ and from Eqs. (\ref{b491}), (%
\ref{b492}) we find
\begin{equation}
\label{b493}E_{s0}\approx \frac{8\pi \hbar R_{*}^2\mu ^{3/2}}{\sqrt{3m}G}%
\sqrt{a_{12}/\sqrt{a_{11}a_{22}}-1},
\end{equation}
$$
E_{sa}\approx 1.008\sqrt{\frac{R_{*}}R}E_{s0}\ll E_{s0}.
$$
{\bf The interface energy of the asymmetric state is negligible}. Close to
the point of phase transition the energy difference between the asymmetric
and symmetric configuration is given by
\begin{equation}
\label{b499}E_a-E_0=\frac 12\mu N_1\left[ (1-\kappa )^2-\frac{4\sqrt{6}\xi }{%
R_{*}}\sqrt{a_{12}/\sqrt{a_{11}a_{22}}-1}\right] \text{,}
\end{equation}
where $\xi =\hbar /\sqrt{2m\mu }$.

In the absence of the interface tension ($\xi \rightarrow 0$) the asymmetric
configuration has greater energy no matter what is the value of $\kappa $ in
comparison with $1$. However, the interface tension makes the asymmetric
phase preferable at some $\kappa $ and results in symmetric-asymmetric phase
transition. The condition $E_a=E_0$ describes a line of global instability
of the system
\begin{equation}
\label{b4991}\frac{a_{11}}{a_{22}}=1-4\left( \frac{\sqrt{6}\xi }{R_{*}}%
\right) ^{1/2}(a_{12}/\sqrt{a_{11}a_{22}}-1)^{1/4}\text{.}
\end{equation}
The soft mode condition gives the line of local stability
\begin{equation}
\label{b4992}\frac{a_{11}}{a_{22}}=1-\frac{4\sqrt{6}\xi }{3R_{*}}(a_{12}/%
\sqrt{a_{11}a_{22}}-1)^{1/2}\text{.}
\end{equation}

In Fig. 9 we plot the phase diagram that shows different stability regions
of two component condensates. In estimates we take $\xi /R_{*}=0.01$. When $%
a_{12}<\sqrt{a_{11}a_{22}}$ the homogeneous binary mixture is a stable
state. Otherwise the two components are phase separated. In the later case
for small ratio $a_{11}/a_{22}$ the state $(1,2)$ with the first component
being inside and the second outside is the only stable configuration. If the
ratio $a_{11}/a_{22}$ increases the configuration becomes globally unstable
when we cross the left (solid) curve. However, the system is locally stable
since the configuration corresponds to a local minima of energy. Further
increase of $a_{11}/a_{22}$ crosses the line of local stability (another
solid line) and the system undergoes a phase transition into a new stable
asymmetric state. If initially the system is prepared in the $(2,1)$ state
then with decreasing the ratio $a_{11}/a_{22}$ the configuration first
becomes globally unstable when we cross the right dotted line and locally
unstable when $a_{11}/a_{22}$ is close enough to $1$.

%%%%%%%%%%%%%%%%%%%% Fig 9 %%%%%%%%%%%%%%%%%%%%%%%%%%%%%
\begin{figure}
\bigskip
\centerline{\epsfxsize=0.48\textwidth\epsfysize=0.48\textwidth
\epsfbox{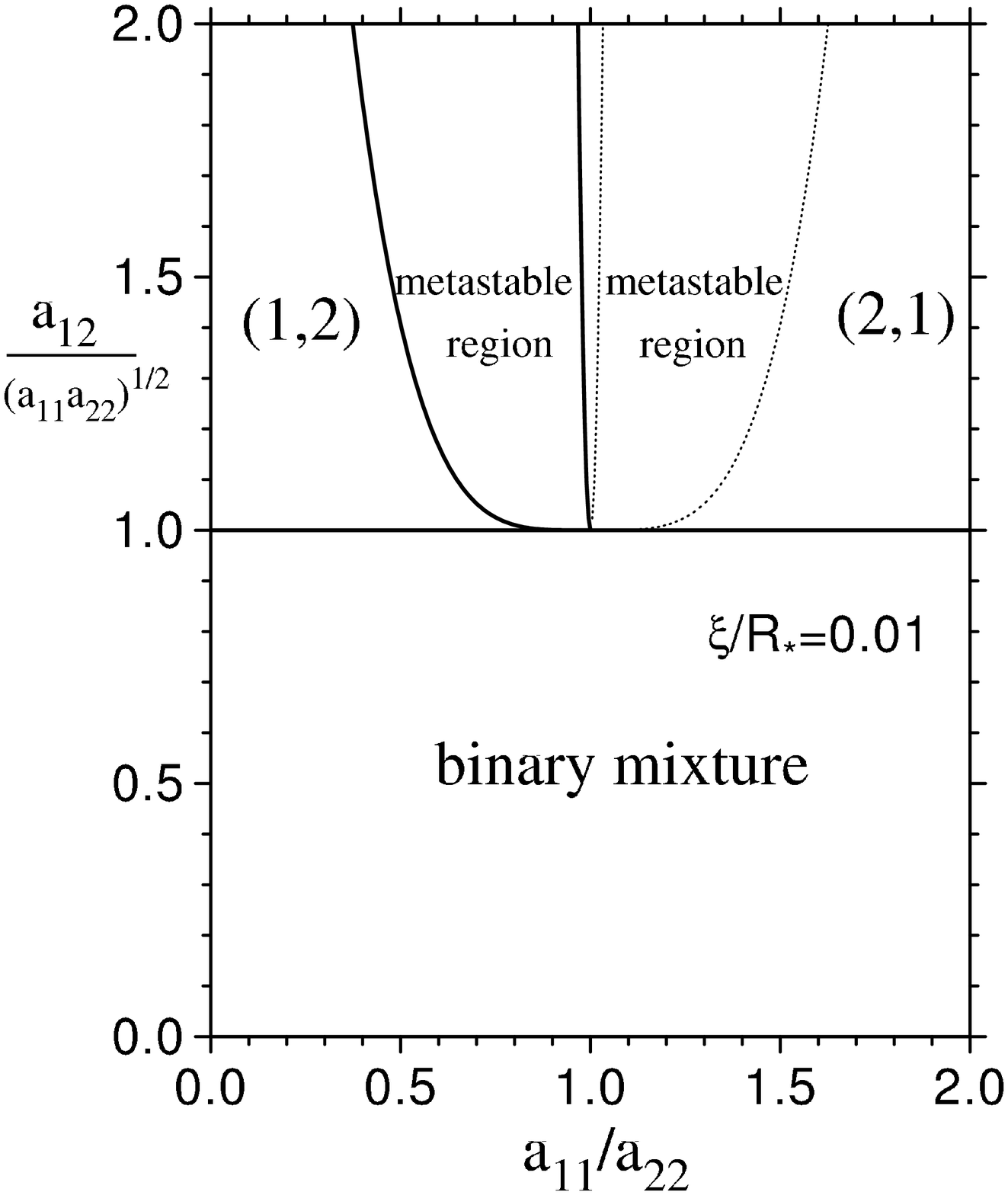}}

\label{fig9}
\end{figure}
%%%%%%%%%%%%%%%%%%%%%%%%%%%%%%%%%%%%%%%%%%%%%%%%%%%%%%%%%

{\small Fig. 9. Phase diagram of the two component condensate in coordinates
displaying relative interaction strength between bosons.}

\section{Energy as a function of droplet displacement}

The question about the system's global stability is related to the energy of
the state. To further explore the metastability of the system here we
estimate the energy of the system for a displaced inner droplet. We assume
the inner droplet has a spherical shape with a radius $r_0$ and the center
of the sphere is displaced a distance $z_0$ from the trap center. The radius
$r_0$ depends on $z_0$. The condensate densities are given by the same TF
expressions (\ref{d1}), (\ref{d2}), however, the position of the interface
is now different. We assume the number of particles of each species are
fixed:
\begin{equation}
\label{b1}N_1=\frac{4\pi }{3G_{11}}\left[ \mu _1r_0^3-\frac{3m\omega
_0^2r_0^5}{10}-\frac{m\omega _0^2r_0^3z_0^2}2\right] ,
\end{equation}
\begin{equation}
\label{b2}N_2=\frac{4\pi }3\frac 1{G_{22}}\left[ \frac 25\mu _2R_2^3-\mu
_2r_0^3+\frac{3m\omega _0^2r_0^5}{10}+\frac{m\omega _0^2r_0^3z_0^2}2\right]
.
\end{equation}
For fixed numbers of particles, the chemical potentials are functions of the
positions of the droplets and are not constants. In the TF limit we find for
the energy of the system%
$$
E=\frac{\mu _1N_1}2+\frac{\mu _2N_2}2+\pi m\omega _0^2\left[ \left( \frac{%
\mu _1}{G_{11}}-\frac{\mu _2}{G_{22}}\right) \left( \frac{r_0^5}5+\frac{%
r_0^3z_0^2}3\right) +\frac{m\omega _0^2}2\left( \frac 1{G_{22}}-\frac 1{%
G_{11}}\right) \left[ \frac{r_0^7}7+\frac{2r_0^5z_0^2}3+\frac{r_0^3z_0^4}3%
\right] +\right.
$$
\begin{equation}
\label{b5}\left. +\frac{2^{7/2}\mu _2^{7/2}}{35G_{22}m^{5/2}\omega _0^5}%
\right] +E_s,
\end{equation}
where $E_s$ is the interface energy. The energy (\ref{b5}) depends on four
parameters: $\mu _1$, $\mu _2$, $z_0$, $r_0$. However, only two parameters
are independent because there are two restrictions imposed by particle
conservation (\ref{b1}), (\ref{b2}). For simplicity let us consider the path
for which $\mu _2=const$. Then only one parameter is independent. If we
choose $r_0$ as an independent variable then the energy as a function of $%
r_0 $ can be expressed in an explicit form%
$$
E=const+\frac{2\pi (\mu _{10}-\mu _2)^2R_{*}^6}{3G_{11}r_0^3}+
$$
\begin{equation}
\label{b9}+4\pi \mu _2^2\left( \frac 1{G_{22}}-\frac 1{G_{11}}\right) \left[
\frac{R_{*}^6}{6r_0^3}+\frac{2r_0^2(r_0^3-R_{*}^3)}{15R_2^2}-\frac{R_{*}^8}{%
5R_2^2r_0^3}+\frac{3R_{*}^{10}}{50R_2^4r_0^3}+\frac{2r_0^2R_{*}^5}{25R_2^4}-%
\frac{12r_0^7}{175R_2^4}\right] +E_s,
\end{equation}
where $\mu _2=m\omega _0^2R_2^2/2$, $\mu _{10}$ is the chemical potential at
$z_0=0$ and $R_{*}$ is the droplet radius at $z_0=0$. The dependence of the
droplet radius on displacement $z_0$ is described by Eq. (\ref{b2}).

\subsection{Interface energy}

Here we estimate the interface energy as a function of the droplet
displacement $z_0$. According to Ref. \cite{Timm98} the interface tension is
given by
\begin{equation}
\label{s1t}\sigma =\frac 4{\sqrt{3}}\sqrt{(\xi _1^2+\xi _2^2)\left[ a_{12}/%
\sqrt{a_{11}a_{22}}-1\right] }P,
\end{equation}
where $\xi _i$ represents the single condensate coherence length $\xi
_i=\hbar /\sqrt{2m_iG_{ii}n_i}$, the pressure $P\approx G_{ii}n_i^2/2$ and
the condensate densities $n_i$ are estimated near the interface. The
interface tension $\sigma $ depends on the local density and, hence, changes
along the interface. To calculate $\sigma $ we assume that the system is
close to the point of phase transition, that is $G_{11}\approx G_{22\text{ }%
}=G$, $R_1\approx R_2=R$, $\mu _1\approx \mu _2=\mu $ and near the interface
$n_1\approx n_2=n$. As a result,

\begin{equation}
\label{s2t}\sigma \approx \frac{2\hbar \sqrt{G}n^{3/2}}{\sqrt{3}\sqrt{m}}%
\sqrt{a_{12}/\sqrt{a_{11}a_{22}}-1}
\end{equation}
and we find for the surface energy
\begin{equation}
\label{s4}E_s=\frac{4\pi \hbar \mu ^{3/2}R^2r_0}{5\sqrt{3m}Gz_0}\sqrt{a_{12}/%
\sqrt{a_{11}a_{22}}-1}\left[ \left( 1-\frac{(r_0-z_0)^2}{R^2}\right)
^{5/2}-\left( 1-\frac{(r_0+z_0)^2}{R^2}\right) ^{5/2}\right] .
\end{equation}

\subsection{Energy for small inner droplet}

In the limit $r_0\ll R$ ($N_1\ll N_2$) Eq. (\ref{b2}) results in a simple
relation between the droplet radius $r_0$ and its displacement $z_0$: $%
r_0^3\approx \mu _2R_{*}^3/(\mu _2-m\omega _0^2z_0^2/2)$. Then near the
point of phase transition the energy of the system is given by
\begin{equation}
\label{s8}E\approx const+N_1\mu \left[ (1-\kappa )\frac{z_0^2}{R^2}+b\left(
1-\frac{z_0^2}{R^2}\right) ^{5/6}\right] ,
\end{equation}
where%
$$
b=\frac{2\sqrt{6}\xi }{R_{*}}\sqrt{a_{12}/\sqrt{a_{11}a_{22}}-1}=\frac{%
3\sigma _0}{2PR_{*}}.
$$
The first term in Eq. (\ref{s8}) corresponds to energy contribution due to
difference between $a_{11}$ and $a_{22}$, it increases with $z_0$ when $%
\kappa <1$. The second term is the contribution from the interface energy,
it decreases with increasing $z_0$ because surface tension drops when the
droplet moves in the region with smaller density. In Fig. 10 we plot the
energy (\ref{s8}) as a function of displacement of the inner droplet $z_0$.
The energy is shown for $b=0.3$ and different $\kappa $. Far from the point
of phase transition the energy monotonically increases with displacement of
the droplet, in this case the first term dominates. When $\kappa $ goes
closer to $1$ the first term in Eq. (\ref{s8}) becomes small and the
interface contribution changes the energy behavior. The energy becomes
nonmonotonic function of displacement. The value of $\kappa $ when the final
and initial energies are equal is the onset of global instability. However,
the system is still locally stable since position of the droplet at the
center corresponds to a local minima of energy. In this region an energy
barrier prevents the droplet to move outside. Further increase of $\kappa $
results in disappearance of the energy barrier and initial configuration
becomes both locally and globally unstable.

%%%%%%%%%%%%%%%%%%%% Fig 10 %%%%%%%%%%%%%%%%%%%%%%%%%%%%%
\begin{figure}
\bigskip
\centerline{\epsfxsize=0.48\textwidth\epsfysize=0.52\textwidth
\epsfbox{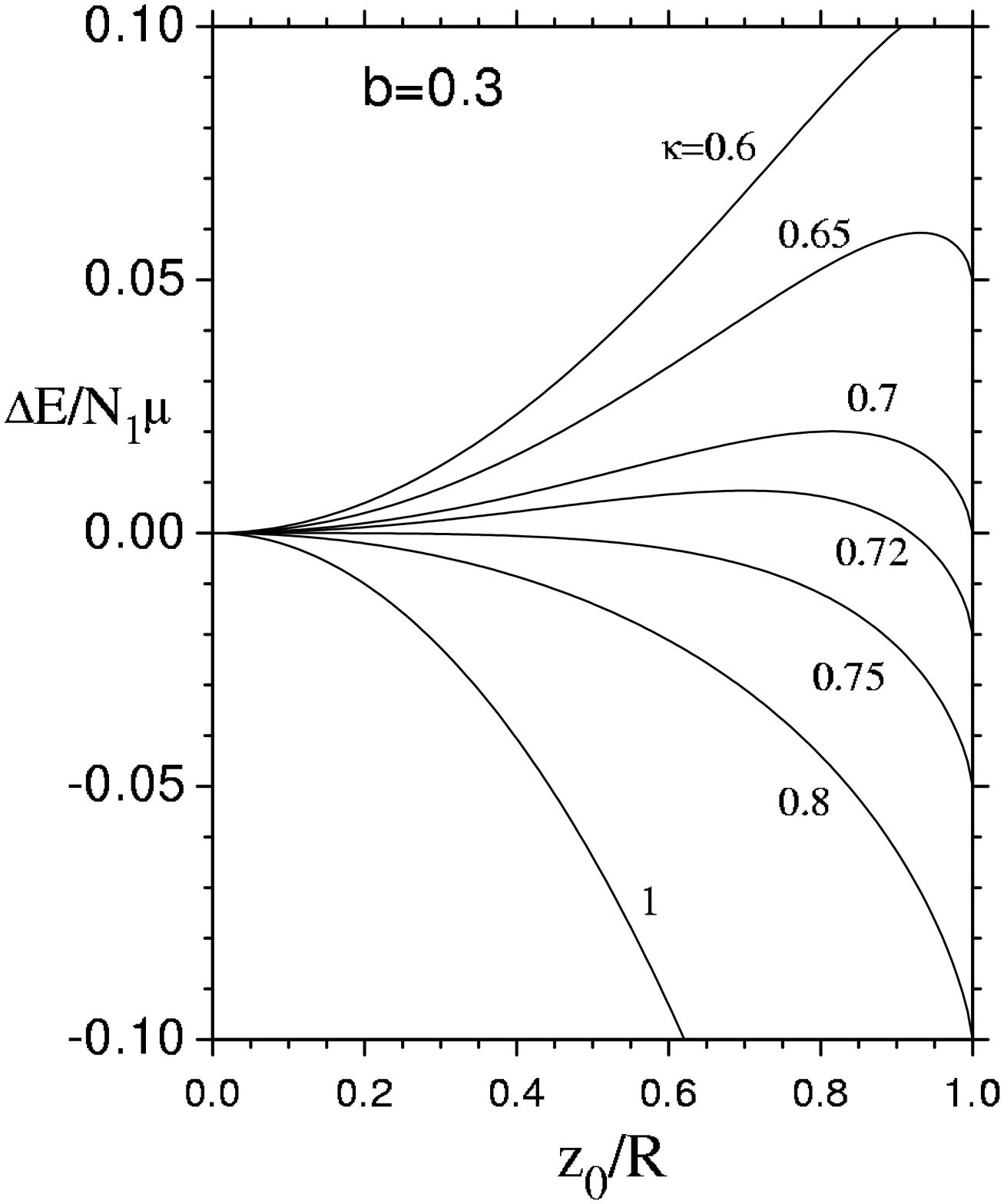}}

\label{fig10}
\end{figure}
%%%%%%%%%%%%%%%%%%%%%%%%%%%%%%%%%%%%%%%%%%%%%%%%%%%%%%%%%

{\small Fig. 10. Energy of the system as a function of displacement of
the inner droplet for fixed $b=0.3$ and different $\kappa$.}

\vspace{0.5cm}

Near the point of local instability the energy barrier is $\Delta E\approx
18N_1\mu (\kappa _s-\kappa )^2/5b$, where $\kappa _s=1-5b/6$. The barrier
becomes of the order of $\mu $ when $\kappa _s-\kappa \approx \sqrt{5b/18N_1}
$. If $N_1\sim 10^6$, $b\sim 0.1$ we obtain $\kappa _s-\kappa \sim 10^{-4}$,
while $1-\kappa _s\sim 0.1$.

\section{Normal modes for different trap frequencies and particle masses of
the two species}

In this section we generalize our results to the case $\omega _1\neq \omega
_2$, $m_1\neq m_2$ which is useful for future experiments where the trap
frequencies and particles can differ for each component. For simplicity we
omit the interface tension and consider spherical trapping potentials. The
generalized boundary condition for the velocity potential $\Phi $ reads
\begin{equation}
\label{f1}m_1\omega _1^2R_{*}\frac{\partial \Phi _1}{\partial r}-m_1\omega
^2\Phi _1=\kappa m_2\omega _2^2R_{*}\frac{\partial \Phi _2}{\partial r}%
-\kappa m_2\omega ^2\Phi _2,
\end{equation}
where $\kappa =\sqrt{G_{11}/G_{22}}=\sqrt{a_{11}m_2/a_{22}m_1}$. The
boundary condition for $\partial \Phi _1/\partial r$ remains the same as Eq.
(\ref{s11}). Now the phase transition is controlled by a new dimensionless
parameter%
$$
\tilde \kappa =\frac{m_2\omega _2^2}{m_1\omega _1^2}\sqrt{\frac{G_{11}}{%
G_{22}}}=\frac{m_2^{3/2}\omega _2^2}{m_1^{3/2}\omega _1^2}\sqrt{\frac{a_{11}%
}{a_{22}}}.
$$
Instead of (\ref{e6}), the solutions for the inner and outer condensates
have the form
\begin{equation}
\label{f4}\Phi _1=C_1r^lF\left( \alpha _1,\beta _1,l+3/2,\frac{r^2}{R_1^2}%
\right) ,\quad \Phi _2=C_2r^lF\left( \alpha _2,\beta _2,1,1-\frac{r^2}{R_2^2}%
\right) ,
\end{equation}
where $R_i^2=2\mu _i/m_i\omega _i^2$ ($i=1,2$) and%
$$
\alpha _i=\frac 12\left[ l+\frac 32-\sqrt{l^2+l+\frac 94+\frac{2\omega ^2}{%
\omega _i^2}}\right] ,\quad \beta _i=\frac 12\left[ l+\frac 32+\sqrt{l^2+l+%
\frac 94+\frac{2\omega ^2}{\omega _i^2}}\right] .
$$
Using Eq. (\ref{f4}) and the boundary conditions we obtain an equation for
the normal mode frequencies
\begin{equation}
\label{f5}\omega ^2=\frac{\omega _1^2(1-\tilde \kappa )\left[
l(l+3/2)+(l-\omega ^2/\omega _1^2)\lambda xs_1(\omega ,x)\right] \left[
(l-\omega ^2/\omega _2^2)xs_2(\omega ,x)-l\right] }{\left[ \tilde \kappa
(l-\omega ^2/\omega _1^2)\lambda xs_1(\omega ,x)\omega _1^2/\omega
_2^2+(l+3/2)\left( l(\tilde \kappa \omega _1^2/\omega _2^2-1)+(l-\omega
^2/\omega _2^2)xs_2(\omega ,x)\right) \right] },
\end{equation}
where%
$$
\lambda =\frac{m_1\omega _1^2\mu _2}{m_2\omega _2^2\mu _1},\quad x=\frac{%
R_{*}^2}{R_2^2},\quad s_1(\omega ,x)=\frac{F\left( \alpha _1+1,\beta
_1+1,l+5/2,\lambda x\right) }{F\left( \alpha _1,\beta _1,l+3/2,\lambda
x\right) },\quad s_2(\omega ,x)=\frac{F\left( \alpha _2+1,\beta
_2+1,2,1-x\right) }{F\left( \alpha _2,\beta _2,1,1-x\right) }.
$$
In terms of $\tilde \kappa $ and $\lambda $ the position of the interface is
given by%
$$
R_{*}=R_2\sqrt{\frac{1-\tilde \kappa \lambda }{\lambda (1-\tilde \kappa )}}.
$$

In the region $|1-\tilde \kappa |\ll 1\,$one can take $\omega =0$ in the
right side of Eq. (\ref{f5}) and put $\tilde \kappa $,$\lambda \approx 1$ in
the multiple. As a result, we find
\begin{equation}
\label{f7}\omega ^2\approx \omega _1^2(1-\tilde \kappa )f(l,x),
\end{equation}
where%
$$
f(l,x)=\frac{l\left( l+3/2+xs_1(0,x)\right) \left( xs_2(0,x)-1\right) }{%
\left[ xs_1(0,x)\omega _1^2/\omega _2^2+(l+3/2)\left( \omega _1^2/\omega
_2^2-1+xs_2(0,x)\right) \right] }>0.
$$
Eq. (\ref{f7}) is a generalization of Eq. (\ref{s19}) for the low frequency
modes. The point of instability is now determined by the condition $\tilde
\kappa =1$, or $a_{11}m_2^3\omega _2^4=a_{22}m_1^3\omega _1^4$. $\tilde
\kappa $ can be changed not only by changing the relative interaction
strengths, but also by changing the ratio of the trapping frequencies.
Furthermore it is proportional to a {\bf higher} (4th) power of the ratio $%
\omega _1/\omega _2$ as compared to $a_{11}/a_{22}$ and thus will be very
sensitive to the frequency change. Near the instability point $\omega
\propto \sqrt{1-\tilde \kappa }$ and becomes imaginary when $\tilde \kappa
>1 $.

In summary, we study the normal modes and the stability of the phase
separated two component BECs and predict a new symmetric-asymmetric
transition in such a system. We show that the normal mode frequencies differ
from those for the one component condensate and new branches appear. We
found normal modes with low frequencies which are analogous to gravitational
waves at the interface between two immiscible fluids. Under the change of
the relative interaction strength $a_{11}/a_{22}$ or the ratio $\omega
_1/\omega _2$ the frequencies of those modes go to zero and become
imaginary. Imaginary frequencies mean instability and the system undergoes a
quantum phase transition into a new stable configuration. Such transition
can be observed experimentally by manipulating the interaction strength via
Feschbach resonances or by changing the trap frequencies. Interface tension
shifts the normal mode frequencies and results in stability of the
asymmetric state in some parameter region. Another interesting possibility
is to study tunneling by turning on a microwave field which allows direct
exchange of the two components and has been done in many JILA experiments
with two hyperfine states of $^{87}$Rb. If we suddenly switch the species,
one can study a decay of new unstable configuration or tunneling directly,
instead of trying to tune the ratio $a_{11}/a_{22}$.

We are grateful to A. Fetter for valuable remarks. This work was supported
by NASA, Grant No. NAG8-1427.


\begin{references}
\bibitem{Myat97}  C.J. Myatt, E.A. Burt, R.W. Ghrist, E.A. Cornell, and C.E.
Wieman, Phys. Rev. Lett. {\bf 78}, 586 (1997).

\bibitem{Hall98}  D.S. Hall, M.R. Matthews, J.R. Ensher, C.E. Wieman, and
E.A. Cornell, Phys. Rev. Lett. {\bf 81}, 1539 (1998).

\bibitem{Stam98}  D.M. Stamper-Kurn, M.R. Andrews, A.P. Chikkatur, S.
Inouye, H.-J. Miesner, J. Stenger, and W. Ketterle, Phys. Rev. Lett. {\bf 80}%
, 2027 (1998).

\bibitem{Sten98}  J. Stenger, S. Inouye, D.M. Stamper-Kurn, H.-J. Miesner,
A.P. Chikkatur, and W. Ketterle, Nature (London) {\bf 396}, 345 (1998).

\bibitem{Mies99}  H.-J. Miesner, D.~M. Stamper-Kurn, J.~Stenger, S.~Inouye,
A.~P. Chikkatur, and W.~Ketterle, Phys. Rev. Lett., {\bf 82}, 2228, (1999).

\bibitem{Stam99b}  D.~M. Stamper-Kurn, H.-J. Miesner, A.~P. Chikkatur,
S.~Inouye, J.~Stenger, and W.~Ketterle, Phys. Rev. Lett., {\bf 83}, 661,
(1999).

\bibitem{Matt98}  M.R. Matthews, D.S. Hall, D.S. Jin, J.R. Ensher, C.E.
Wieman, E.A. Cornell, F. Dalfovo, C. Minniti, and S. Stringari, Phys. Rev.
Lett. {\bf 81}, 243 (1998).

\bibitem{Hall98b}  D.~S. Hall, M.~R. Matthews, C.~E. Wieman, and E.~A.
Cornell, Phys. Rev. Lett., {\bf 81}, 1543, (1998).

\bibitem{Matt99a}  M.~R. Matthews, B.~P. Anderson, P.~C. Haljan, D.~S. Hall,
C.~E. Wieman, and E.~A. Cornell, Phys. Rev. Lett., {\bf 83}, 2498, (1999).

\bibitem{Khal57}  I.M. Khalatnikov, Zh. \'Eksp. Teor. Fiz. {\bf 32}, 653
(1957) [Sov. Phys. JETP {\bf 5}, 542 (1957)].

\bibitem{Sigg80}  E.D. Siggia, and A.E. Ruckenstein, Phys. Rev. Lett. {\bf 44%
}, 1423 (1980).

\bibitem{Ho96}  T.-L. Ho, and V.B. Shenoy, Phys. Rev. Lett. {\bf 77}, 3276
(1996).

\bibitem{Law97}  C. K. Law, H. Pu, N.P. Bigelow, and J.H. Eberly, Phys. Rev.
Lett. {\bf 79}, 3105 (1997).

\bibitem{Esry97}  B.D. Esry, C.H. Greene, J.P. Burke, and J.L. Bohn, Phys.
Rev. Lett. {\bf 78}, 3594 (1997).

\bibitem{Gold97}  E.V. Goldstein, and P. Meystre, Phys. Rev. A {\bf 55},
2935 (1997).

\bibitem{Ohbe98}  P. \"Ohberg, and S. Stenholm, Phys. Rev. A. {\bf 57}, 1272
(1998).

\bibitem{Pu98}  H. Pu, and N.P. Bigelow, Phys. Rev. Lett. {\bf 80}, 1130
(1998).

\bibitem{Ao98}  P. Ao, and S. T. Chui, Phys. Rev. A {\bf 58}, 4836 (1998).

\bibitem{Timm98}  E. Timmermans, Phys. Rev. Lett. {\bf 81}, 5718 (1998).

\bibitem{Chui99}  S.~T. Chui and P.~Ao, Phys. Rev. A, {\bf 59}, 1473, (1999).

\bibitem{Chui3}  S.~T. Chui, P.~Ao, and B.~Tanatar, J. Phys. Soc. Jpn., {\bf %
15}, 142, {1999}.

\bibitem{Busc97}  T. Busch, J.I. Cirac, V.M. P\'erez-Garc\'ia, and P.
Zoller, Phys. Rev. A {\bf 56}, 2978 (1997).

\bibitem{Grah98}  R. Graham, and D. Walls, Phys. Rev. A {\bf 57}, 484 (1998).

\bibitem{Esry98}  B.D. Esry, and C.H. Greene, Phys. Rev. A {\bf 57}, 1265
(1998).

\bibitem{Gord98}  D. Gordon, and C.M. Savage, Phys. Rev. A {\bf 58}, 1440
(1998).

\bibitem{Maze02}  I.E. Mazets, Phys. Rev. A {\bf 65}, 033618 (2002).

\bibitem{Pu98a}  H. Pu, and N.P. Bigelow, Phys. Rev. Lett. {\bf 80}, 1134
(1998).

\bibitem{Burk97}  J.P. Burke, Jr., J.L. Bohn, B.D. Esry, and C.H. Greene,
Phys. Rev. A {\bf 55}, R2511 (1997).

\bibitem{Chui02}  S.T. Chui, V.N. Ryzhov, and E.E. Tareyeva, Pis'm. Zh.
\'Eksp. Teor. Fiz. {\bf 75}, 279 (2002) [JETP Letters {\bf 75}, 233 (2002)].

\bibitem{Fett96}  A.L. Fetter, Phys. Rev. A {\bf 53}, 4245 (1996).

\bibitem{Stri96}  S. Stringari, Phys. Rev. Lett. {\bf 77}, 2360 (1996).

\bibitem{Land88a}  L. D. Landau, and E. M. Lifshits, {\it Hydrodynamics},
Moscow, Nauka, (1988), Sec. 12.

\bibitem{qsd}  P. Ao and S. T. Chui, J. Phys. B: At. Mol. Opt. Phys. {\bf 33}%
, 535 (2000).

\bibitem{Land88}  L.D. Landau, and E.M. Lifshits, {\it Hydrodynamics},
Moscow, Nauka, (1988), Sec. 62.
\end{references}
\end{document}